\documentclass[
twocolumn, showpacs,preprintnumbers,amsmath,amssymb]{revtex4}

\usepackage{graphicx}
\usepackage{dcolumn}
\usepackage{bm}
\usepackage{epsfig}
\usepackage{natbib}

\begin{document}


\title{The combined effect of chemical and electrical synapses in small
  Hindmarsh-Rose neural networks on synchronisation and on the rate
  of information}

\author{M. S. Baptista$^1$, F. M. Moukam Kakmeni$^2$, C. Grebogi$^1$}
\affiliation{$^1$ Institute for Complex Systems and Mathematical Biology,
King's College, University of Aberdeen, AB24 3UE Aberdeen, United
Kingdom \\
$^2$ Laboratory of Research on Advanced Materials and Nonlinear Science (LaRAMaNS),
Department of Physics, Faculty of sciences, University of Buea,
P. O. Box 63 Buea, Cameroon}

\date{\today}

\begin{abstract}
  In this work we studied the combined action of chemical and
  electrical synapses in small networks of Hindmarsh-Rose (HR) neurons
  on the synchronous behaviour and on the rate of information produced
  (per time unit) by the networks. We show that if the chemical
  synapse is excitatory, the larger the chemical synapse strength used
  the smaller the electrical synapse strength needed to achieve complete
  synchronisation, and for moderate synaptic strengths one should
  expect to find desynchronous behaviour. Otherwise, if the chemical
  synapse is inhibitory, the larger the chemical synapse strength used
  the larger the electrical synapse strength needed to achieve complete
  synchronisation, and for moderate synaptic strengths one should
  expect to find synchronous behaviours.  Finally, we show how to
  calculate semi-analytically an upper bound for the rate of
  information produced per time unit (Kolmogorov-Sinai entropy) in
  larger networks. As an application, we show that this upper bound is
  linearly proportional to the number of neurons in a network whose
  neurons are highly connected.
\end{abstract}


\maketitle
\section{Introduction}
\noindent

Intercellular communication is one of the most important
characteristics of all animal species because it makes the many
components of such complex systems operate together.  Among the
many types of intercellular communication, we are interested in the
communication among brain cells, the neurons, that exchange
information mediated by chemical and electrical synapses
\cite{sudhof}.

The uncovering of the essence of behaviour and perception in animals
and human beings is one of the main challenges in brain research.
While the behaviour is believed to be linked to the way neurons are
connected (the topology of the neural network and the physical
connections among the neurons), the perception is believed to be
linked to synchronisation. This comes from the binding hypothesis
\cite{malsburg1}, which states that synchronisation functionally binds
neural networks coding the same feature or objects.  This hypothesis
raised one of the most important contemporary debates in neurobiology
\cite{pareti} because desynchronisation seems to play an important
role in perception as well.  The binding hypothesis is mainly
supported by the belief that a convenient environment for neurons to
exchange information appears when they become more synchronous.

Despite the explosive growth in the field of complex networks, it is
still unclear for which conditions synchronisation implies information
transmission and it is still unclear which topology favours the
flowing of information. Additionally, most of the models being
currently studied in complex networks consider networks whose nodes
(such as neurons) are either linearly or non-linearly connected. But,
recent works have shown that neurons that were believed to make only
non-linear (chemical) synapses make also simultaneously linear
(electrical) synapses
\cite{Gibson,Connors,Galarreta1999,Hestrin,Galarreta2001}. To make the
scenario even more complicated, neurons connect chemically in an
excitatory and/or an inhibitory way. In this work, we aim to study the
relationship between synchronisation and information transmission in
such neural networks, whose neurons are simultaneously connected by
chemical and electrical synapses.

The electrical synapse is the result of the potential difference
between the neurons and causes an immediate physiological response of
the latter one, linearly proportional to the potential difference. The
chemical synapse is mediated by the exchange of neurotransmitters from
the pre to the postsynaptic neuron and can only be released once the
presynaptic neuron membrane achieves a certain action potential.  The
chemical interaction is described by a nonlinear function
\cite{Greengard}.

While the electrical synapses between neurons is localised in the
neuron cell and therefore it is a local connection, the chemical
synapse is in the neuron axon and is therefore mainly responsible for
the non-local nature of the synapses.

Chemical synapses can be inhibitory and excitatory. When an inhibitory
neuron spikes (the pre-synaptic neuron), a neuron connected to it
(the post-synaptic neuron) is prevented from spiking. As shown in Ref.
\cite{vree}, inhibition promotes synchronisation.  When an excitatory
neuron spikes, it induces the post-synaptic neuron to spike.
Several types of synchronisation were found in networks of chaotic
neurons coupled with only electrical synapses. One can have complete
synchronisation, generalised synchronisation and phase
synchronisation, the latter appearing for small synapse strength
\cite{baptista_PRE2008}.  Complete synchrony strongly depends on the
network structure and the number of cells. In networks of chemically
coupled neurons \cite{Baptista1}, the net input a neuron receives from
synaptic neurons emitting synchronised spikes is proportional to the
number of connected units. Hence, for chemical synapses, if all the
nodes in the network have the same degree, synchronisation will be
enhanced; if different nodes have different degrees, synchronisation
will be hampered \cite{Cosenza}. In fact, Ref. \cite{Hasler} has shown
analytically that the stability of the completely
synchronous state in such networks only depends on the number of
signals each neuron receives, independent of all other details of the
network topology.

The most obvious possible role of electrical synapses within networks
of inhibitory neurons is to couple the membrane potential of connected
cells, leading to an increase in the probability of synchronised
action potentials. This synchronous firing could coordinate the
activity of other cortical cell populations. For example, it has been
reported that the introduction of electrical synapses among GABAergic
neurons that are also chemically connected can promote oscillatory
rhythmic activity \cite{Hestrin}.  These possibilities have been
addressed experimentally by several investigators and have been
reviewed recently\cite{Galarreta2001,Connors,Bennett}.

Motivated by these observations and also by the fact that the
behaviour of micro-circuitry in the cerebral cortex is not well
understood, we analyse the combined effect of these two types of
synapses on the stability of the synchronous behaviour and on the
information transmission in small neural networks. In order to deal
with this problem analytically we consider idealistic networks,
composed of equal neurons with mutual connections of equal strengths
(see Sec. \ref{network}). A basic assumption characterising most of
the early works on synchronisation in neural networks is that, by
adding a relatively small amount of electrical synapse to the
inhibitory synapse, one can increase the degree of synchronisation far
more than a much larger increase in inhibitory conductance
\cite{Kopell,Pfeuty}.

Our results agree with this finding in the sense that for larger
inhibitory synaptic strengths complete synchronisation can only be
achieved if the electrical synapse strength is larger than a certain
amount. But in contrast, we found that for moderate inhibitory synaptic
strengths, the larger the chemical synapse strength is the larger the
electrical synapse strength needs to be to achieve complete synchronisation.
Additionally, we introduce in this work analytical approaches to
understand when complete synchronisation should be expected to be found and
what is the relation of that with the amount of information produced
by the network.

Information is an important concept \cite{shannon}. It measures how
much uncertainty one has about an event before it happens. It is a
measure of how complex a system is. Very complicated and higher
dimensional systems might be actually very predictable, and as a
consequence the content of information of such a system might be very
limited. But measuring the amount of information is something
difficult to accomplish.  Normally, there is always some bias or error
on the calculation of it \cite{paninski}, and one has to rely on
alternative approaches.  Measuring the Shannon entropy of a chaotic
trajectory is extremely difficult because one has to calculate an
integral of the probability density of a fractal chaotic set. But for
chaotic systems that have absolutely continuous conditional measures,
one can calculate Shannon's entropy per unit of time, a quantity known
as Kolmogorov-Sinai (KS) entropy \cite{kolmogorov}, by summing all the
positive Lyapunov exponents \cite{LS}. A system that has absolutely
continuous conditional measures is a system whose trajectory
continuously distribute along unstable directions. More precisely,
systems whose trajectories continuously distribute along unstable
manifolds at points that have positive probability measure.  These
systems form a large class of nonuniformly hyperbolic systems
\cite{young}: the H\'enon family; H\'enon-like attractor arising from
homoclinic bifurcations; strange attractors arising from Hopf
Bifurcations (e.g.  R\"ossler oscillator); some classes of mechanical
models with periodic forcing. The result in Ref.  \cite{LS} extends a
previous result by Pesin \cite{pesin} that demonstrated that for
hyperbolic maps, the KS entropy is equal to the sum of the positive
Lyapunov exponents. We are not aware of any rigorous result proving
the equivalence of the KS entropy and the sum of Lyapunov exponent for
the Hindmarsh-Rose neural model neither to a network constructed with
them. But the chaotic attractors arising in this neuron model are
similar to the ones appearing from Homoclinic bifurcations.
Additionally, for two coupled neurons, we show in Sec.
\ref{combined_information} (using the non-rigorous methods described
in Appendix \ref{apendice1}) that a lower bound estimation of the KS
entropy is indeed close to the sum of all the positive Lyapunov
exponents.  Despite the lack of a rigorous proof, we will assume that
the results in Refs. \cite{LS,young} apply in here in the sense that
the sum of the positive Lyapunov exponents provide a good estimation
for the KS entropy.

The KS entropy for chaotic networks has another important meaning. It
provides one the so called network capacity \cite{baptista_PRE2008},
the maximal amount of information that all the neurons in the network
can simultaneously process (per unit of time). A network that produces
information at a higher rate is more unpredictable and more complex.
Arguably, the network capacity is an upper bound for the amount of
information that the network is capable of processing from external
stimuli. In Ref.  \cite{baptista_PRE2008} we discuss a situation were
that is indeed the case.

To understand the scope of this paper and the methods used, we first justify
the chosen network topologies in Sec.  \ref{network}. Then, in
Sec. \ref{network_dynamic}, we describe the dynamical system of our
network and derive the variational equations of it in the eigenmode
form, a necessary analytical tool in order to be able to study the
onset of complete synchronisation (CS) and to calculate the rate of
information produced by the network. Complete synchronisation happens
when the trajectories of all neurons are equal. 

Our main results can be summarised as in the
following:
\begin{itemize}
\item We show (Secs.  \ref{estabilidade} and \ref{rescaling}) how one
  can calculate the synaptic strengths (chemical and electrical)
  necessary for a network of $N$ neurons to achieve complete
  synchronisation when one knows the strengths for which two mutually
  coupled neurons become completely synchronous.  
\item We show numerically (Sec.  \ref{combined_synchronous}) parameter
  space diagrams indicating the electrical and chemical synapse
  strengths responsible to make complete synchronisation to appear in
  different networks. The analytical derivation from Sec.
  \ref{rescaling} are found to be sufficiently accurate. There are
  two scenarios for the appearance of complete synchronisation for
  inhibitory networks. If the chemical synapse strength is small, the
  larger the chemical synapse strength used the larger the electrical
  synapse strength needed to be to achieve complete synchronisation.
  Otherwise, if the chemical synapse strength is large, complete
  synchronisation appears if the electrical synapse strength is lager
  than a certain value.  In excitatory networks both synapses work in
  a constructive way to promote complete synchronisation: the larger
  the chemical synapse strength is the smaller the electrical synapse
  strength needs to be to achieve complete synchronisation.
\item We show (Secs. \ref{combined_information}) that the sum of the
  positive Lyapunov exponents provides a good estimation for the KS
  entropy. Additionally, we show that there are optimal ranges of
  values for the chemical and electrical strengths for which the
  amount of information is large. 
\item If complete synchronisation is absent, we 
show (Sec.  \ref{EX_IN}) that while in
inhibitory networks one can typically expect to find high levels of
synchronous behaviour, in excitatory networks one is likely to expect
desynchronous behaviour.  
\item We calculate (Sec.  \ref{bounds}) an upper bound for the rate of
  information produced per time unit (Kolmogorov-Sinai entropy) by
  larger networks using the rate at which information is produced by
  two mutually coupled neurons.
\end{itemize}

\section{The topology of the studied networks}\label{network}

In order to consider the combined action of these two different types
of synapses, we need to consider in our theoretical approach
idealistic networks, constructed by nodes possessing equal dynamics
and particular coupling topologies such that a synchronisation
manifold exists and CS is possible. If we had studied networks whose
neurons were exclusively connected by electrical means, we could have
considered networks with arbitrary topologies. On the other hand, if
we had studied networks whose neurons are exclusively connected by
chemical means, we would have considered networks whose neurons
receive the same number of chemical connections. These conditions
are the same ones being usually made to study complete synchronisation
in complex networks \cite{Pecora,Hasler}.

In order to analytically study networks formed by neurons that make
simultaneously chemical and electrical connections, we have not only
to assume that the neurons have equal dynamics and that every neuron
receives the same number of chemical connections coming from other
neurons, but also that the Laplacian matrix for the electrical
synapses (that provides topology of the electrical connections) and
the Laplacian matrix for the chemical synapses commute, as we clarify
later in this paper.  Naturally, there is a large number of Laplacian
matrices that commute. In this work we construct networks that are
biologically plausible. Since the electrical connection is local, we
consider that neurons connect electrically only to their nearest
neighbours. Since neurons connected chemically make a large number of
connections (of the order of 1000), it is reasonable to consider that
for small networks the neurons that are chemically connected are fully
connected, i.e., every neuron connects to all the other neurons.
Notice however that while reciprocal connections are commonly found in
electrically coupled neurons, that is not typical for chemically
connected neurons.

Since our small networks are composed of no more than 8 neurons, we
make an abstract assumption and admit another possible type of network
in which neurons that are connected electrically can also make
non-local connections, allowing them to become fully connected to the
other neurons. Notice, however, that our theoretical approach remains
valid for larger networks that admit a synchronisation manifold.

\section{The networks of coupled neurons and master stability
analysis}\label{network_dynamic} \noindent

The dynamics of the Hindmarsh-Rose (HR) model for neurons is described
by

\begin{eqnarray}
\dot{p}&=& q- a p^3 + b p^2 -n + I_{ext} \nonumber \\
\dot{q}&=& c - d p^2 - q \\
\dot{n}&=& r[s(p - p_0)-n] \nonumber
\end{eqnarray}
\noindent
where $p$ is the membrane potential, $q$ is associated with the fast
current, $Na^+$ or $K^+$, and $n$ with the slow current, for example,
$Ca^{2+}$. The parameters are defined as $a=1, b=3, c=1, d=5, s=4,
r=0.005, p_0=-1.60$ and $I_{ext}=3.2$ where the system exhibits a
multi-time-scale chaotic behaviour characterised as spike-bursting.

The dynamics of a neural networks of $N$ neurons connected
simultaneously by electrical (a linear coupling) and chemical (a
non-linear coupling) synapses is described by
\begin{eqnarray}\label{neurons02}
\dot{p_i}&=& q_i-a p_i^3 +b p_i^2 -n_i + I_{ext}\nonumber\\
&&- g_n(p_i-V_{syn})\sum_{j=1}^{N}\mathbf{C}_{ij}S(p_j)
+ g_l\sum_{j=1}^{N}\mathbf{G}_{ij}\mathbf{H}(p_j)\nonumber\\
\dot{q_i}&=& c - d p_i^2 - q_i \\
\dot{n_i}&=& r[s(p_i - p_0)-n_i] \nonumber
\end{eqnarray}
\noindent
$(i,j)=1,\ldots,N$, where $N$ is the number of neurons. 

In this work we consider that $\mathbf{H}(p_i)=p_i$. But we preserve
the function $\mathbf{H}(p_i)$ in our remaining analytical derivation
to maintain generality.

The chemical synapse function is modelled by the sigmoidal function
\begin{equation}
S(p_j)=\displaystyle\frac{1}{1+e^{-\lambda(p_j -\Theta_{syn})}},
\label{sp}
\end{equation}
\noindent
with $\Theta_{syn}=-0.25$, $\lambda=10$ and $V_{syn}=2.0$ for
excitatory and $V_{syn}=-2.0$ for inhibitory.  For the chosen
parameters and all the networks that we have worked $|p_i|<2$ and 
the term $(p_i-V_{syn})$ is always negative for excitatory networks
and positive for inhibitory networks.  If two neurons are connected
under an inhibitory (excitatory) synapse then, when the presynaptic
neuron spikes, it induces the postsynaptic neuron not to spike (to
spike).  

The matrix $\mathbf{G}_{ij}$ describes the way neurons are
electrically connected. It is a Laplacian matrix and therefore $\sum_j
\mathbf{G}_{ij} = 0$. The matrix $\mathbf{C}_{ij}$ describes the way
neurons are chemically connected and it is an adjacent matrix,
therefore $\sum_j \mathbf{C}_{ij} = k$, for all $i$. For both
matrices, a positive off-diagonal term placed in the line $i$ and
column $j$ means that neuron $i$ perturbs neuron $j$ with an intensity
given by $g_l \mathbf{G}_{ij}$ (or by $g_n \mathbf{C}_{ij}$). Since
the diagonal elements of the adjacent matrix are zero, $k$ represents
the number of connections that neuron $i$ receives from all the other
neurons $j$ in the network.  This is a necessary condition for the
existence of the synchronous solution \cite{Hasler} by the subspace $P
= P_1 = P_2 = . . = P_N, P_i = (p_i, q_i, n_i)$.

Under these assumptions and, as previously explained, we consider
networks with three topologies: {\bf topology I}, when all the neurons
are mutually fully (all-to-all) connected with chemical synapses and
mutually diffusively (nearest neighbours) connected with electrical
synapses; {\bf topology II}, when all the neurons are mutually fully
connected with chemical synapses and mutually fully connected with
electrical synapses; {\bf topology III}, when all the neurons are
mutually diffusively (nearest neighbours) connected with chemical and
electrical synapses. We consider networks with 2, 4 and 8 neurons.  By
nearest neighbours, we consider that the neurons are forming a closed
ring.

The synchronous solutions $P = (p, q, n)$ take the form
\begin{eqnarray}\label{neurons04}
\dot{p}&=& q-a p^3 +b p^2 -n + I_{ext}
- g_n k(p-V_{syn})S(p)\nonumber\\
\dot{q}&=& c - d p^2 - q \\
\dot{n}&=& r[s(p - p_0)-n] \nonumber
\end{eqnarray}
The variational equation of the network in (\ref{neurons02}) [calculated
around the synchronisation manifold (\ref{neurons04})] is given by
\begin{eqnarray}\label{neurons06}
\dot{\delta p_i}&=&\delta q_i-3 a p_i^2\delta p_i + 2b p_i\delta p_i - \delta n_i \nonumber\\
&&- g_n(p_i-V_{syn})S^{'}(p)\left(k \delta p_i+\sum_{j=1}^{N}\mathbf{\tilde{G}}_{ij}\delta p_j\right)\nonumber\\
&& -k g_n S(p)\delta p_i + g_l\sum_{j=1}^{N}\mathbf{G}_{ij}D \mathbf{H}(p)\delta p_i\nonumber\\
\dot{\delta q_i}&=&  2d \delta p_i- \delta q_i \\
\dot{\delta n_i}&=& r(s\delta p_i-\delta n_i) \nonumber
\end{eqnarray}
\noindent
The matrix $\mathbf{C}_{ij}$ has been transformed to a Laplacian
matrix by $\mathbf{\tilde{G}}=\mathbf{C}_{ij}-k \mathbb{I}$. $D
\mathbf{H}(p)$ represents the derivative of $\mathbf{H}$ with respect
to $p$, which in this work equals 1.

The term $S^{\prime}(p)$ refers to the spatial derivative $\frac{d
  S(p)}{dp}$ and equals
\begin{equation}
S^{\prime}(p)=\frac{\lambda \exp{^{-\lambda(p - \Theta_{syn})}}}{[1+ \exp{^{-\lambda(p - \Theta_{syn})}}]^2}.
\label{Sprime}
\end{equation}
\noindent
Notice that if $S(p)=1$ (what happens for $p>>\Theta_{syn}$), then
$S^{\prime}(p)=0$ and if $S(p)=0$ ($p << \Theta_{syn}$), then
$S^{\prime}(p)=0$. $S^{\prime}(p)$ is not zero when the value of
$S(p)$ changes from 1 to 0 (and vice-versa) and $p \approxeq
\Theta_{syn}$.

Equation (\ref{neurons06}) is referred to as the variational equation
and is often the starting point for determining whether the
synchronisation manifold is stable. This equation is rather
complicated since, given arbitrary synapses $g_n$ and $g_l$, it can
become quite higher dimensional. Also the coupling matrices
$\mathbf{G}$ and $\mathbf{\tilde{G}}$ can be arbitrary making the
situation to become even more complicated.  However, assuming that
whenever there is a chemical synapse (and $g_n > 0$), the matrices
$\mathbf{G}$ and $\mathbf{\tilde{G}}$ commute, then the problem can be
simplified by noticing that the arbitrary state $\delta X$ (where
$\delta X=(\delta p_i, \delta q_i, \delta n_i)$ is the deviation of
the $i$th vector state from the synchronisation manifold) can be
written as $\delta X=\sum_{i=1}^N \textbf{v}_i \bigotimes
\kappa_i(t)$, with $\kappa_i(t)=(\eta_i,\psi_i,\varphi_i)$.  The
$\textbf{v}_i$ be the eigenvector and $\gamma_i$ and
$\tilde{\gamma}_i$ the corresponding eigenvalues for the matrices
$\mathbf{G}$ and $\mathbf{\tilde{G}}$ respectively. So, if that is the
case, by applying $\textbf{v}_j^T(t)$ (with $\textbf{v}_j^T(t)\cdot
\textbf{v}_i=\delta_{ij}\:\: \text{where} \:\: \delta_{ij} \:\:
\text{is the Kronecker delta}$), to the left (right) side of each term
in Eq. (\ref{neurons06}) one finally obtains the following set of N
variational equations in the eigenmode
\begin{eqnarray}
\dot{\eta_j}&=&(2b p-3 a p^2)\eta_j - \varphi_j+ \psi_j-\Gamma(p)\eta_j\nonumber\\
\dot{\psi_j}&=&  2d \eta_j- \psi_j \nonumber \\
\dot{\varphi_j}&=& r(s\eta_j-\varphi_j) \label{neurons07}\\
j&=&1, 2, 3,  ...N\nonumber
\end{eqnarray}
where the term $\Gamma(p)$ is given by
\begin{equation}
\Gamma(p)= k g_n S(p) - g_n(V_{syn}-p)S^{'}(p)\left(k
  +\tilde{\gamma}_j\right) - g_l\gamma_j
\label{gama}
\end{equation} 
\noindent
in which $\gamma_j$ (with
$\gamma_1$=0, and $\gamma_j<$0, $j \geq 2$) are the eigenvalues of
$\mathbf{G}$ and $\tilde{\gamma_j}$ are the eigenvalues of
$\mathbf{\tilde{G}}$.  The eigenvalues $\gamma_j$ are negative because
the off-diagonal elements of $\mathbf{G}$ are positive.

For networks with $N=2$ we have that $|\gamma_2|=2$ and $k=1$, meaning
that the neurons are connected in an all-to-all fashion. For networks
with $N=4$, if the neurons are connected in an all-to-all fashion, we
have that $|\gamma_2|=4$ and $k=3$ or if the neurons are connected with
their nearest neighbours we have that $|\gamma_2|=2$ and $k=2$.  For
$N=8$, $|\gamma_2|=8$ and $k=7$ (all-to-all) and $|\gamma_2|=0.585786402$
and $k=2$ (nearest-neighbour). These values are placed in Table
\ref{tabela1} for further reference.

\begin{table}
\caption{Values of $\gamma_2$ in absolute value and $k$ for the considered networks.}
\label{tabela1}
\begin{tabular}{|c|c|c|}\hline
 & all-to-all & nearest-neighbour \\ \hline $N=2$ & $\gamma_2=2$, $k$=1
 & $\gamma_2=2$, $k$=1 \\ $N=4$ & $\gamma_2=4$, $k$=3 & $\gamma_2=2$,
 $k$=2 \\ $N=8$ & $\gamma_2=8$, $k$=7 & $\gamma_2=0.585786402$, $k$=2
 \\ \hline
\end{tabular}
\end{table}

The previous equations are integrated using
the 4th-order Range-Kutta method with a step size of 0.001. The
calculations of the Lyapunov exponents are performed considering a
time interval of 600 [sufficient for a neuron to produce approximately
600 spikes ($p>0$)].  We discard a transient time of 300,
corresponding to 300,000 integrations.

\section{Stability analysis}\label{estabilidade}

The stability of the synchronisation manifold can be seen from the
perspective of control \cite{Hasler,Femat,Moukam,Bowong} by imagining
that the term $\Gamma(p)$ stabilises Eq. (\ref{neurons07}) at the
origin. This term can be interpreted as the main gain of a feedback
control law $u(t)=\Gamma(p)\eta_j$ such that $\eta_j$ (resp.  $\psi_j$
and $\varphi_j$ ) tends to $0$ as $t$ tends to infinity. In fact, the
controlling force $u(t)=\Gamma(p)\eta_j$ could be designed with no
previous knowledge of the system under consideration assuming that it
has a parametric dependence.  A drawback of such a general control
approach is that it leads to non-feedback control strategy, which
have not guaranteed stability margins. More robust approaches for
determining the structural stability of the synchronisation manifold
of systems whose equations of motion are partially unknown have been
recently developed \cite{Femat, Moukam, Bowong}.

In this work, however, we determine the stability of the
synchronisation manifold from the master stability analysis of Refs.
\cite{Hasler,Pecora}. A necessary condition for the linear stability
of the synchronised state is that all Lyapunov exponents associated
with $\gamma_j$ and/or $\tilde{\gamma_j}$ for each $j=2, 3, ..., N$
(the directions transverse to the synchronisation manifold) are
negative.  This criterion is a necessary condition for complete
synchronisation only locally, i.e. close to the synchronisation
manifold.

\section{Rescaling of Eqs. (\ref{neurons04}) and (\ref{neurons07})}\label{rescaling}

When working with networks formed by nodes possessing equal dynamical
rules, we wish to predict the behaviour of a large network from the
behaviour of two coupled nodes. That can always be done whenever the
equations of motion of the network can be rescaled into the form of
the equations describing the two coupled nodes.  That means that,
given that two mutually coupled neurons completely synchronise for
the electrical and chemical synapse strengths $g_l^*(N=2)$ and
$g_n^*(N=2)$, respectively, then it is possible to calculate the
synapse strengths $g_l^*(N)$ and $g_n^*(N)$ for which a network
composed by $N$ nodes completely synchronises.

In order to rescale the equations for the synchronisation manifold and
for its stability, Eqs. (\ref{neurons04}) and (\ref{neurons07}),
respectively, we need to preserve the form of these equations as we
consider different networks.

Concerning Eq. (\ref{neurons07}), we need to show under which
conditions it is possible to have $\Gamma(p,N=2)=\Gamma(p,N)$, where
$\Gamma$ is the term responsible to make the stability of the
synchronisation manifold to depend among other things on the topology
of the network and on the coupling function $S(p)$.

Notice that $S(p)$ assumes for most of the time either the value 0 or
1. For some short time interval $S(p)$ changes its value from 0 to 1
(and vice-versa) and at this time $S^{\prime}(p)$ is different from
zero [see Eqs. (\ref{sp}) and (\ref{Sprime})]. For that reason we will
treat $S^{\prime}(p)$ as a small perturbation in our further
calculations and will ignore it, most of the times. That leave us with
two relevant terms in both Eqs.  (\ref{neurons04}) and
(\ref{neurons07}) that need to be taken into consideration in our
rescaling analyses.  These terms are $g_l \gamma_j$ and $k g_n S(p)$.
While the first term comes from the electrical synapse, the second
term comes from the chemical synapse.

The first term depends on the eigenvalues of $\mathbf{G}_{ij}$ (which
varies according to the number of nodes and the topology of the
network) and on the synapse strength $g_l$. If this term assumes a
particular value for a given network, for another network one can
suitably vary $g_l$ in order for the whole term to assume this same
value in the other network. So, the term $g_l \gamma_j$ can always be
rescaled by finding an appropriate value of $g_l$.

The rescaling of the second term, $k g_n S(p)$ is more
complicated because it depends on the trajectory $(p)$ of the
attractor.  Naturally, we wish to find a proper rescaling for the
function $S(p)$, which implies that the attractors appearing as
solutions on the synchronisation manifold should present some kind of
invariant property.

\begin{figure}[!h]
  \centerline{\hbox{\psfig{file=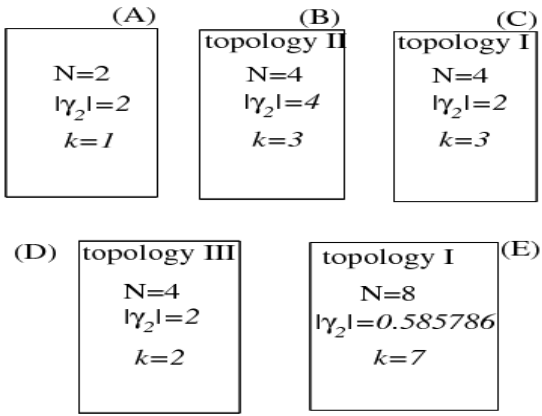,height=6.0cm,width=8.0cm,
       viewport=0 -20 400 410}}} \caption{The topology of the
   networks considered in Figs. \ref{S_P_inib54_58}, \ref{S_P_67_71},
   \ref{PS_PECORA_inib78_82} and \ref{PS_PECORA_24_28} and the values
   of $N$, $|\gamma_2|$ and $k$.}
\label{mostra_valores}
\end{figure}

\begin{figure}[!h]
  \centerline{\hbox{\psfig{file=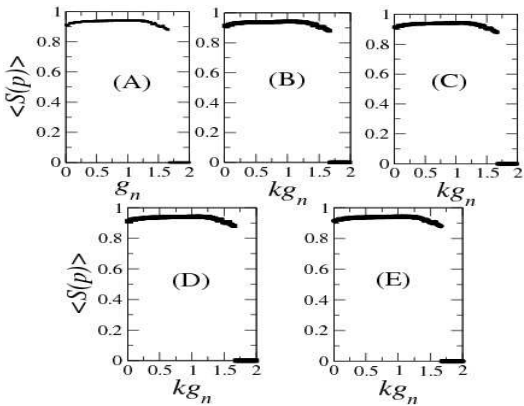,height=6.0cm,width=8.0cm,
        viewport=0 0 500 510}}} \caption{(A-E) The value of $\langle
    S(p) \rangle$ with respect to a rescaled chemical synapse strength
    $kg_n$ for excitatory networks with a configuration shown in Figs.
    \ref{mostra_valores}(A-E). Initial conditions of the neurons are
    set to be equal (and $g_l$=0).}
\label{S_P_inib54_58}
\end{figure}

\begin{figure}[!h]
 \centerline{\hbox{\psfig{file=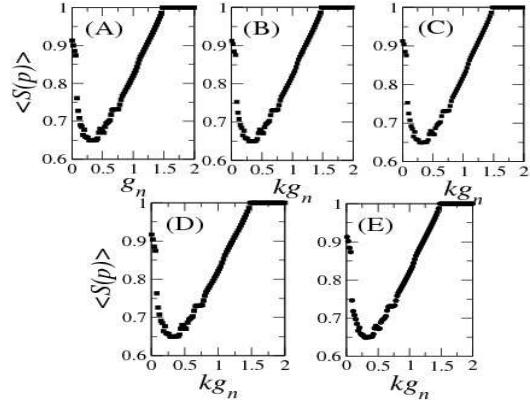,height=6.0cm,width=8.0cm,
        viewport=0 0 500 510}}} \caption{(A-E) The value of $\langle
    S(p) \rangle$ with respect to a rescaled chemical synapse
    strength $kg_n$ for inhibitory networks with a
    configuration shown in Figs. \ref{mostra_valores}(A-E). Initial conditions of the neurons are
    set to be equal (and $g_l$=0).}
\label{S_P_67_71}
\end{figure}

In order to find such an invariant property, we study the time average
$\langle S(p) \rangle$ of the function $S(p)$ for attractors appearing
as solutions of Eq. (\ref{neurons04}) for 5 network topologies. In
Fig.  \ref{mostra_valores} we show in the boxes (A-E) the values of
$N, $$|\gamma_2|$, $k$ and the type of topology considered in the
networks of Figs. \ref{S_P_inib54_58}, \ref{S_P_67_71},
\ref{PS_PECORA_inib78_82}, and \ref{PS_PECORA_24_28}.

The result for excitatory networks can be seen in Fig.
\ref{S_P_inib54_58}(A-E), which shows this value as a function of $k
g_n$. Apart from some small differences, the function $\langle
S(p)\rangle$ remains invariant for the different networks considered.
We identify two relevant values for $\langle S(p)\rangle$. Either
$\langle S(p)\rangle \approxeq 0.9$,
for $g_n < g_n^{(c)}$ or $\langle S(p)\rangle=0$, for $g_n \geq
g_n^{(c)}$. $g_n^{(c)} \approx 1.67$. 

We also find an invariant curve of $\langle S(p)\rangle$ for
inhibitory networks. In Fig. \ref{S_P_67_71}(A-E) we show this curve
for the same networks of Fig. \ref{S_P_inib54_58}. For these networks,
we define $g_n^{(c)} \approx 1.5$ as the value of $g_n$ for which the
curve of $\langle S(p)\rangle$ reaches its maximum. In the considered
inhibitory networks, $\langle S(p)\rangle=1$ is a consequence of the
fact that the neurons loose their chaotic behaviour and become a
stable limit cycle.  Notice that the value of $\langle S(p) \rangle$
does not depend on the value of the electrical synapse strength $g_l$.
This is due to the fact that $g_l$ is not present in the equations for
the synchronisation manifold [Eq.  (\ref{neurons04})].

Let us rescale Eq. (\ref{neurons04}). First notice that the average
$\langle (p-V_{syn}) \rangle$ has the same invariant properties of the
average $\langle S(p) \rangle$. Then, we assume that both $S(p)$ and
$(p-V_{syn})$ make small oscillations around their average value. That
implies that $S(p) (p-V_{syn}) \approxeq \langle S(p) (p-V_{syn})
\rangle$. From Figs.  \ref{S_P_inib54_58} and \ref{S_P_67_71} we have
that the average $\langle S(p,N) \rangle$ can be written as a function
of $g_n(N)$, as well as $\langle (p-V_{syn}) \rangle$. Therefore, we
can write $\langle S(p) (p-V_{syn}) \rangle$ as a function of
$g_n(N)$. It is clear that the value of this average obtained for
$g_n(N=2)$ should be approximately equal to the value obtained for
$kg_n(N)$, and so this average function can be rescaled by $kg_n(N)
\approxeq g_n(N=2)$.  Therefore, Eq.  (\ref{neurons04}) describing a
large network can be rescaled into this same equation describing two
mutually coupled neurons by

\begin{eqnarray}
\label{neurons081}
g_n(N)=\displaystyle\frac{g_n(N=2)}{k}
\end{eqnarray}

Now, we need to show that it is also possible to do the same to Eq.
(\ref{neurons07}), the equation responsible for the stability of the
synchronous solution.

Assuming again that $S(p)$ make small oscillations around its average value
allows us to write $\Gamma(p,N)$ as a function of $\langle S(p)
\rangle$ as in $\Gamma(p,N) \cong kg_n(N) \langle S(p,N) \rangle -
g_l(N) \gamma_j$. Notice from Figs.  \ref{S_P_inib54_58} and
\ref{S_P_67_71} that the average $\langle S(p,N) \rangle$ can be
written as a function of $g_n(N)$. In order to rescale Eq.
(\ref{neurons07}), describing a network of $N$ nodes in terms of a
network of 2 nodes, we need to have that $\Gamma(p,N)=\Gamma(p,N=2)$
leading to
\begin{eqnarray}
k g_n(N) \langle S[g_n(N)] \rangle - \gamma_2 g_l(N) &=& \nonumber \\
g_n(N=2) \langle S[g_n(N=2)] \rangle + 2 g_l(N) \label{neurons13}
\end{eqnarray}
\noindent
where we have considered only the second largest eigenvalue
$\gamma_2$, the one responsible for the stability of the
synchronisation manifold; we have ignored terms that appear
together with $S^{\prime}$ in $\Gamma$.

We make now a reasonable hypothesis that if a stable synchronous
solutions for Eq.  (\ref{neurons04}) exists for $g_n(N=2)=g_n^*(N=2)$
(for a two mutually coupled neurons), then this same stable
synchronous solution exists for $k g_n^*(N)$ (for a network composed
by $N$ neurons mutually connected). This hypothesis is constructed
from the observation that equivalent attractors can be found in
different networks if the rescaling in Eq. (\ref{neurons081}) is
employed. We are assuming that if $g_n^*(N=2)$ represents the chemical
synapse strength for which complete synchronisation appears in two
mutually coupled neurons, then complete synchronisation would appear
in a network of $N$ nodes if
\begin{eqnarray}
\label{neurons004}
g^*_n(N)=\displaystyle\frac{g^*_n(N=2)}{k}
\end{eqnarray}

If the previous hypothesis is satisfied, i.e. Eq. (\ref{neurons004})
is satisfied, we see from Figs.  \ref{S_P_inib54_58} and
\ref{S_P_67_71} that $\langle S[g_n(N)] \rangle \approxeq \langle
S[g_n(N=2)] \rangle$ and assuming that these two averages are equal,
then Eq. (\ref{neurons13}) takes us to 
\begin{eqnarray}
\label{neurons09}
g^*_l(N)=\displaystyle\frac{2 g^*_l{(N=2)}}{|\gamma_2(N)|}
\end{eqnarray}
\noindent
where $g^*_l(N)$ represents the electrical synapse strength for which
complete synchronisation occurs in a network composed by $N$ neurons.

In the following, we analyse two special cases of Eq.
(\ref{neurons13}) when the function $S(p)$ is constant and the
previous approximations (expanding $\Gamma$ around its average and
that $\langle S[g_n(N)] \rangle = \langle S[g_n(N=2)] \rangle$) to
arrive to Eqs. (\ref{neurons004}) and (\ref{neurons09}) are exact.

\subsection{Rescaling in excitatory networks ($V_{syn}=2.0$)}

{\it Case 1}: A large chemical synapse strength, $ k g_n(N)
>g_n^{(c)}$, with $g_n^{(c)} \approxeq $1.67, makes for {\bf all} the time
$p < \Theta$, leading to $S(p) = 0$ and $S^{\prime}(p)$=0 (see Fig.
\ref{S_P_inib54_58}). The neurons become completely synchronous to a
stable equilibrium point.

\subsection{Rescaling in inhibitory networks ($V_{syn}=-2.0$)}

{\it Case 2}: a large chemical synapse strength, $k g_n(N)>g_n^{(c)}$, 
with $g_n^{(c)} \approx 1.50$, makes for {\bf all} the time $p >
\Theta$ and as a consequence $S(p) = 1$ and $S(p)^{\prime}=0$ (see
Fig. \ref{S_P_67_71}).  The neurons become completely synchronous to a
limit cycle.

\section{Combined effect of the chemical and electrical synapses on
  the synchronous behaviour}\label{combined_synchronous}

The analytical derivations done in the previous section are
approximations, except for some special values of the synaptic
strengths (case 1 and 2). However, as we show in this section, our
calculations provide a good estimation of what to expect from
parameter spaces of larger networks when the parameter space of two
mutually coupled neurons is known.  The parameter space is constructed
by considering the synapses $(g_l,g_n)$ and they identify the regions
where the state of complete synchronisation is stable.

The stability is determined from Eqs. (\ref{neurons07}), by verifying
whether there are no lyapunov exponents associated with transversal
directions to the synchronisation manifold. These exponents are
numerically obtained, without any approximation.

\begin{figure}[!h]
  \centerline{\hbox{\psfig{file=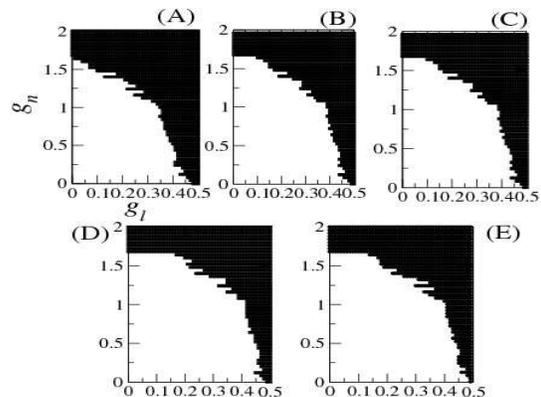,height=6.0cm,width=8.0cm,
        viewport=0 0 500 510}}} \caption{Excitatory networks. Black
    points represent values of the synapse strengths for which all
    transversal conditional exponents are negative. In (B-E) the
    horizontal axis represent $g_l(N)|\gamma_2(N)|/2$ and the vertical
    axis $kg_n$. Initial conditions of the neurons are set to be equal.}
\label{PS_PECORA_inib78_82}
\end{figure}

\begin{figure}[!h]
 \centerline{\hbox{\psfig{file=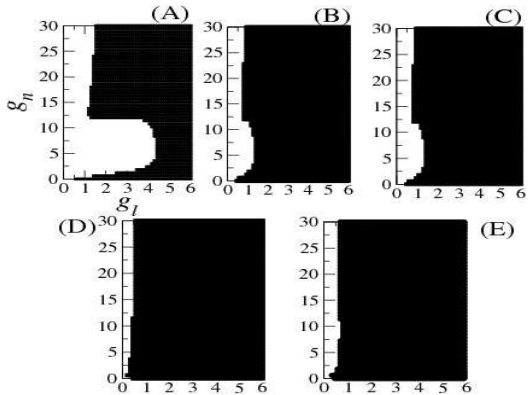,height=6.0cm,width=8.0cm,
        viewport=0 0 500 510}}} \caption{Inhibitory networks. Black
    points represent values of the synapse strengths for which all
    transversal conditional exponents are negative.  In (B-E) the
    horizontal axis represent $g_l(N)|\gamma_2(N)|/2$ and the vertical
    axis $kg_n$. Initial conditions of the neurons are set to be
    equal.}
\label{PS_PECORA_24_28}
\end{figure}

In Fig. \ref{PS_PECORA_inib78_82}, we show in black the synchronous
regions (all transversal conditional exponents are negative) for the
excitatory networks and in Fig.  \ref{PS_PECORA_24_28} the same
network topologies but for inhibitory networks. To simplify the
understanding of these two figures, in Fig.  \ref{mostra_valores} we
show in boxes (A-E) the values of $N$, $|\gamma_2|$, $k$ and the type
of topology considered in the networks of Figs.
\ref{PS_PECORA_inib78_82}(A-E) and \ref{PS_PECORA_24_28}(A-E). The
values of $g_l$ and $g_n$ were rescaled by using Eqs.
(\ref{neurons004}) and (\ref{neurons09}).  As expected, in excitatory
networks our rescaling works very well and roughly in inhibitory
networks. So, the vertical axis of Figs.
\ref{PS_PECORA_inib78_82}(B-E) and \ref{PS_PECORA_24_28}(B-E) show the
quantity $k g_n(N)$ and the horizontal axis of these same figures show
the quantity $\frac{|\gamma_2|g_l(N)}{2}$.

To assist the analysis of the parameter spaces, imagine a curve
$\Sigma$ that is the border between the regions defining parameters
for which the synchronisation manifold is unstable (white regions) and
regions defining parameters for which the synchronisation manifold is
stable (black regions). There are four main characteristics in these
two types (excitatory and inhibitory) of networks concerning the
occurrence of complete synchronisation.

\begin{itemize}
\item[$\bullet$] {\bf In excitatory networks}, the electrical and the
  chemical synapses act in a combined way to foster synchronisation.
  The neurons become completely synchronous to a stable equilibrium
  point. The asynchronous neurons (white regions) are chaotic. The
  curve $\Sigma$ would look like a diagonal line with a negative
  slope. Such a curve could be defined by an equation similar
  to $k g(N) + \gamma_2 g_l \approx C$, $C$ being a function that is
  approximately constant (see Fig.  \ref{PS_PECORA_inib78_82}).
\item[$\bullet$] {\bf In excitatory networks, with $kg_n(N)>$1.67},
  Neurons are completely synchronous to a stable equilibrium
  point (see Fig.
  \ref{PS_PECORA_inib78_82}).
\item[$\bullet$] {\bf In inhibitory networks, with $kg_n(N)<$5}, the
  larger the chemical synapse strength is the larger the electrical
  synapse strength needs to be to achieve complete synchronisation.
  Neurons become completely synchronous to either a limit cycle (large
  chemical synapse strength) or to a chaotic attractor (small chemical
  synapse strength). The curve $\Sigma$ would look like a diagonal
  line with a positive slope. Such a curve could be defined by
  an equation similar to $k g(N) - \gamma_2 g_l \approx C$, $C$ being
  a function that is approximately constant (see Fig.
  \ref{PS_PECORA_24_28}).
\item[$\bullet$] {\bf In inhibitory networks, for large values of $kg_n(N)$},
  complete synchronisation appears for $\gamma_2 g_l > C$ and neurons become completely 
  synchronous to a stable limit cycle, which is unstable if
  $\gamma_2 g_l < C$. The curve $\Sigma$ would look like a straight
  vertical line. Such a curve could be defined by an equation similar
  to $\gamma_2 g_l \approx C$. $C$ being a function that is
  approximately constant (see Fig. \ref{PS_PECORA_24_28}).
\end{itemize}

If the neurons are set with different initial conditions, but
sufficiently close, complete synchronisation is found for similar 
synaptic strengths for which the synchronisation manifold is
stable.

If the neurons are set with sufficiently different initial conditions,
and we construct parameter spaces that represent synaptic strengths
for which CS takes place, we would have obtained parameter spaces with
similar structure as the one observed in Figs.
\ref{PS_PECORA_inib78_82} and \ref{PS_PECORA_24_28}.  However, the
network can become completely synchronous to other synchronous
solutions of Eq. (\ref{neurons04}), different from the synchronous
solutions observed for the parameters used to make Figs.
\ref{PS_PECORA_inib78_82} and \ref{PS_PECORA_24_28}. In other words,
parameter spaces that show CS in networks whose neurons are set with
different initial conditions constructed for the same synaptic
strengths and networks considered in Figs.  \ref{PS_PECORA_inib78_82}
and \ref{PS_PECORA_24_28} would present additional black points in the
white areas of Figs.  \ref{PS_PECORA_inib78_82} and
\ref{PS_PECORA_24_28}.

\section{Combined effect of the chemical and electrical synapses on
  the amount of information}\label{combined_information}

\begin{figure}[!h]
 \centerline{\hbox{\psfig{file=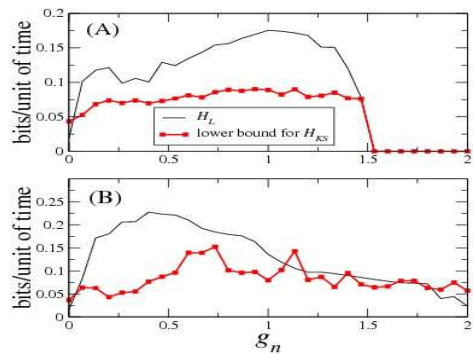,height=6.0cm,width=8.0cm,viewport=0 0 500 510}}}
 \caption{[Color Online] We show the value of the sum of all the
   positive Lyapunov exponents $H_{L}$ in black line and an estimation
   of the lower bound for the KS entropy in filled squares (red line online) for two mutually
   chemically coupled neurons under an excitatory synapse (A) and an
   inhibitory synapse (B), as we vary the chemical synapse strength.
   We consider a constant electrical synapse of strength $g_l$=0.1.
   Initial conditions are not equal.}
\label{estima_entropia_fig02}
\end{figure}

First, we calculate the sum of all the positive Lyapunov exponents of
the attractor obtained from integrating the neural network [Eq.
(\ref{neurons02})] and represent it by $H_L$. The Lyapunov exponents
are calculated from the variational equation of the network in Eq.
(\ref{neurons02}). As previously discussed, it is reasonable to
assume that $H_{L} \approx H_{KS}$, where $H_{KS}$ represents the KS
entropy \cite{kolmogorov}, which measures the amount of information
(Shannon's entropy) produced per time unit. 

In Figs.  \ref{estima_entropia_fig02}(A-B) we show in the thin line
$H_{L}$ for two mutually chemically and electrically coupled neurons
($g_l$=0.1) for excitatory synapse (A) and for inhibitory synapse (B).
To confirm that the sum of the positive Lyapunov exponents have an
entropic meaning for the studied Hindmarsh-Rose neuron model, we have
estimated a lower bound for the KS entropy, represented by the tick
line with filled squares (red online) in Fig.
\ref{estima_entropia_fig02}(A-B).

We see that for both cases, as one increases the synaptic strength,
$H_L$ decreases. For the excitatory case, for $g_n>1.52$, the neurons
trajectories go to an equilibrium point and we obtain $H_L=0$.
If $H_{L}=0$, that means that there are no positive Lyapunov exponents
and therefore no chaos. The maximal value of $H_L$, calculated varying
the synaptic strengths, is almost equal for both types of synapses.
One sees that there is a range of strength values in both figures
within which $H_L$ is large. For example, in (A) $H_L$ is large for
$g_n \in [0.7,1.2]$ and in (B) $H_L$ is large for $g_n \in [0.3,0.7]$.
This was also observed in 3D parameter space diagrams (not shown in
here) that show the value of $H_L$ versus $g_n$ and $g_l$. These
diagrams indicate that there is an optimal range of values for $g_n$
and $g_l$ for which $H_L$ remains large.

The reason we have shown results for two coupled neurons is because
for such a configuration a lower bound estimation of the KS entropy
can be calculated by encoding the trajectory into a binary symbolic
sequence. Since the sequence is binary, this method is only capable of
measuring an information rate that is less or equal than 1bit/symbol
or 1bit/unit of time.  Since that for two coupled neurons, $H_L
<1$bit/unit of time, and assuming that $H_{L}$ is a good estimation
for $H_{KS}$, then the employed method to calculate a lower bound of
the KS entropy is appropriate.  The details of this estimation can be
seen in Appendix \ref{apendice1}.

Notice that in Fig. \ref{estima_entropia_fig02}(A-B) for $g_n
\approx$0 (as well as in (B) for $g_n \approx 2$) the estimations of
$H_{KS}$ are larger than $H_L$. That is the result of a known problem
in the estimation of entropic quantities which prevents the estimation
to be small. The problem arises because the symbolic sequences
considered are not infinitely long for one to realise that there
exists a few or only one symbolic sequence encoding the trajectory.
For example, a long periodic orbit would be encoded by a series of
short symbolic sequences making the estimation of $H_{KS}$ to be
positive instead of zero as it should be.

\section{Synchronisation (and desynchronisation) versus inhibition (and
  excitation) versus Information}\label{EX_IN}

To understand the relation between synchronisation (desynchronisation)
and inhibition (excitability), when {\it complete synchronisation is
  absent} we do the following. But notice that the following results
are based on a conjecture that is currently not demonstrated.

We calculate the Lyapunov exponents along the
synchronisation manifold, which are just the Lyapunov exponents of the
network by assuming that all neurons are completely synchronous.  We call
these exponents conditional Lyapunov exponents and the sum of all the
positive ones is denoted by $H_C$.  There are two ways for calculating
them, either using Eq.  (\ref{neurons06}) or (\ref{neurons07}),
Eq.  (\ref{neurons07}) being simpler because of the dimensionality
of the orthogonal vectors employed to calculate the Lyapunov exponents.
While the use of Eq. (\ref{neurons06}) requires 3N vectors, each one
with dimensionality 3N, the use of Eq.  (\ref{neurons07}) requires N
vectors each one with dimensionality 3.  Additionally, once the
function that relates the conditional exponents of two mutually
coupled neurons with $g_n$ and $g_l$ is known, then one can calculate
this function for all the conditional exponents of larger networks as
long as Eqs. (\ref{neurons04}) and (\ref{neurons07}) can be rescaled.

We can then classify these neural networks into 2 types. The types
UPPER or LOWER. More specifically,

\begin{eqnarray}
H_C(N,g_n,g_l) & > & H_{L}(N,g_n,g_l), \mbox{\ \ \ \ \ UPPER}
\label{conjecture1} \\ H_C(N,g_n,g_l) & < & H_{L}(N,g_n,g_l), \mbox{\
\ \ \ \ LOWER} \label{conjecture4} 
\end{eqnarray}

\noindent To understand what $H_C$ and $H_L$ exactly mean and the
reason for such a classification, notice that the networks here
considered admit a synchronous solution. This synchronous solution
might be unstable (an unstable saddle) and typical initial conditions
depart from the neighbourhood of the synchronous solution and
asymptotically tend towards a stable solution, the chaotic
attractor. This attractor describes a network whose nodes are not
synchronous. In such a situation, the network admits at least two
relevant solutions: a stable desynchronous one (the chaotic attractor)
and an unstable synchronous one (the synchronisation manifold). While
$H_C$ can be associated with the amount of information produced by the
unstable synchronous solution, $H_L$ can be associated with the amount
of information produced by the desynchronous chaotic attractor. If the
complete synchronous state is stable, then, $H_C=H_L$, and the network
in Eq.  (\ref{neurons02}) possesses only one stable synchronous
solution, for typical initial conditions. The nomenclature in
Eqs. (\ref{conjecture1}) and (\ref{conjecture4}) comes from the fact
that if $H_C(N,g_n,g_l) > H_{L}(N,g_n,g_l)$ then, $H_C$ is an upper
bound for $H_L$, otherwise it is a lower bound
\cite{baptista_NJP2008}.

Assume now that the more information a network produces, the more
desynchronisation is observed among pair of neurons
\cite{baptista_NJP2008,baptista_PLOS2008}.  If $H_C(N,g_n,g_l) >
H_{L}(N,g_n,g_l)$ (UPPER), then $H_{L}(N,g_n,g_l)$ is limited. As a
consequence, the production of information in the network is limited
and therefore the level of desynchronisation is small. On the other
hand, if $H_C(N,g_n,g_l) < H_{L}(N,g_n,g_l)$ (LOWER), then
$H_{L}(N,g_n,g_l)$ can be large implying a large level of
desynchronisation.  Another way of understanding the relationship
between synchronisation and information is by using a result from Ref.
\cite{baptista_NJP2008}, which shows that for two coupled maps (but
this result is trivially extended to networks), the largest
transversal conditional exponent, when the maps have a LOWER
character, is larger than this exponent for when they have an UPPER
character.  Since this exponent provides a necessary condition for the
stability of the synchronisation manifold, it can be interpreted as a
measure of the level of desynchronisation in the network.  The larger
this exponent is, the more desynchronous the network is. Therefore,
UPPER networks should have neurons more synchronous than LOWER
networks.

If $H_C(N,g_n,g_l) > H_{L}(N,g_n,g_l)$ (UPPER), the synapse forces the
trajectory to approach the synchronisation manifold and, as a
consequence, there is a high level of synchronisation in the network.
On the other hand, if $H_C(N,g_n,g_l) < H_{L}(N,g_n,g_l)$ (LOWER), the
synapse forces the trajectory to depart from the synchronisation
manifold and, as a consequence, there is a high level of
desynchronisation in the network.

\begin{figure}[!h]
\centerline{\hbox{\psfig{file=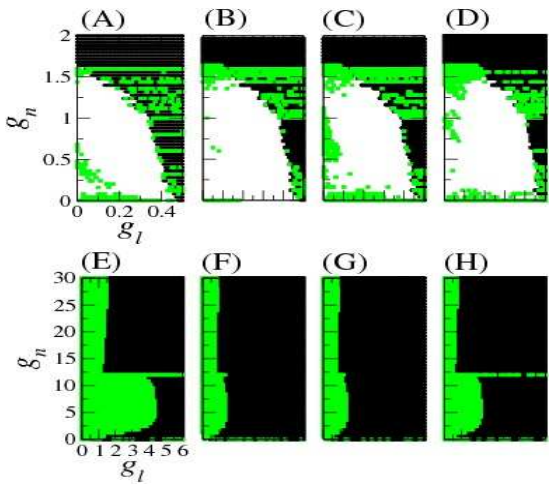,height=7.0cm,width=8.0cm,
        viewport=0 0 430 510}}}
  \caption{[Color online] Gray regions (green online) indicate ($g_n,g_l$) values
    for which $H_C>H_L$ (UPPER) and black regions indicate ($g_n,g_l$)
    values for which the complete synchronisation state is stable, in
    excitatory networks (A-D) and inhibitory networks (E-H). The
    networks considered in (A-D) as well as in (E-H) have the
    parameters shown in Fig.  \ref{mostra_valores}(A-D).  In (B-D) and
    (F-H) the horizontal axis represent $g_l(N)|\gamma_2(N)|/2$ and
    the vertical axis $kg_n$. Gray points (green online) appearing on black regions
    represent synaptic strengths for which in fact one has $H_C =
    H_L$, but numerically we obtain that $H_C = H_L +\epsilon$, with
    $\epsilon$ being a very small positive constant.  }
\label{fig_HKS_HC00_19}
\end{figure}

One can check that in Fig.  \ref{fig_HKS_HC00_19}, which shows as
gray, the parameter regions for which $H_C > H_L$ and as black the
parameter regions for which the synchronisation manifold is stable and
there is complete synchronisation (and therefore, $H_C=H_L$) for
typical initial conditions. Gray points appearing on black regions
represent synaptic strengths for which in fact one has $H_C = H_L$,
but numerically we obtain that $H_C = H_L +\epsilon$, with $\epsilon$
being a very small positive constant.  Typically, neurons coupled via
an excitatory synapse [(A-D)] present a LOWER character while via an
inhibitory synapse [(E-H)] present an UPPER character.

This classification is also important because as it was shown in Ref.
\cite{baptista_NJP2008}, once two coupled neurons are UPPER (or LOWER)
there is always a synaptic strength range for which a large network is
UPPER (or LOWER). And these synaptic strength ranges can be calculated
using the rescalings in Eqs. (\ref{neurons004}) and (\ref{neurons09}).

In Figs. \ref{fig_HKS_HC00_19}(B-C) and \ref{fig_HKS_HC00_19}(F-H), we
show that the UPPER and LOWER character of two mutually coupled
neurons is preserved in networks composed by a number of neurons
larger than 2, if one considers the rescalings of Eqs.
(\ref{neurons004}) and (\ref{neurons09}).  This result is of
fundamental importance, specially for synaptic strengths that promote the
network to have an UPPER character because it allows us to calculate
an upper bound for the KS entropy of larger networks by knowing the
value of $H_C$ for two mutually coupled neurons.  Such a situation
arises for inhibitory networks for a large range of both synaptic
strengths. One finds an UPPER character in excitatory networks for a
small value of the chemical synapse strength.

The electrical synapse favours the neurons to synchronise. As a
consequence, it is expected that networks with neurons connected
exclusively by electrical synapses are of the UPPER type. This can be
checked in all figures for when $g_n \approxeq $0.  

We are currently trying to prove the conjecture in Ref.
\cite{baptista_NJP2008} by studying the relationship between the
stability of unstable periodic orbits \cite{paulo} embedded in the
attractors appearing in complex networks and the stability of the
equilibrium points. All the equilibrium points of a polynomial network
can be calculated by the methods in Refs. \cite{ra1,ra2,ra3}.

\section{Upper bound for the rate of information}\label{bounds}

According to Ruelle \cite{ruelle}, the sum of all the positive Lyapunov
exponents is an upper bound for the Kolmogorov-Sinai entropy
\cite{kolmogorov}.  Therefore, whenever $H_C(N) > H_L(N)$ (UPPER)
it is valid to write that
\begin{equation}
  H_C(N) > H_{KS}(N)
\label{upper}
\end{equation}
\noindent
where $H_{KS}(N)$ denotes the Kolmogorov-Sinai entropy of a network
composed of $N$ neurons.

As we have previously seen, the UPPER character of two mutually
coupled neurons is preserved in the special larger networks here studied. In
addition to this, if the positive conditional exponents of two
mutually coupled neurons are known for a given $g_n$ and $g_l$,
allowing us to calculate $H_C[N=2,g_n(N=2),g_l(N=2)]$, then one can
calculate the positive conditional exponents of a network with $N$
neurons, $H_C[N,g_n(N),g_l(N)]$. In other words, if the ratio of
information production of two mutually coupled neurons that have equal
trajectories, $H_C(N=2)$, is known and the neurons have an UPPER
character, one can calculate the upper bound for the ratio of
information production in larger networks, as long as Eqs.
(\ref{neurons04}) and (\ref{neurons07}) can be rescaled.  Therefore,
in UPPER networks connected simultaneously with electrical and
inhibitory chemical synapses we can always calculate an upper bound
for the rate of information production in terms of this quantity in
two mutually coupled inhibitory neurons.

Consider two mutually coupled neurons. Denote $\lambda_1(N=2,g_n)$ as
the sum for the positive Lyapunov conditional exponents associated
with the synchronisation manifold for a chemical synapse strength
$g_n$ and $\lambda_2(N=2,g_n,g_l)$ as the sum of the positive Lyapunov
exponents associated with the only one transversal direction for a
chemical synapse strength $g_n$ and an electrical synapse strength
$g_l$.  Remind that $\lambda_1$ and $\lambda_2$ are calculated using
Eq.  (\ref{neurons07}) for the index $j=1$ and $j=2$, respectively.

Now, consider a network formed by N neurons. Using similar arguments
than the ones presented in Sec. \ref{rescaling} and based on the conjecture 
proposed in \cite{baptista_NJP2008}, the value of the
synapse strengths $g_l(N),g_n(N)$ for which the exponent
$\lambda_1(N)$ has the same value of $\lambda_1(N=2)$ can be
calculated by
\begin{equation}
g_n(N)=\frac{g_n(N=2)}{k}
\label{eq000}
\end{equation}
\noindent
and the value of the synapse strengths $g_l(N),g_n(N)$ for which the
sum of the positive conditional exponent $\lambda_w(N,g_n,g_l)$ (for
$w \geq 2$) has the same value of $\lambda_2(N=2,g_n,g_l)$ can be
calculated by
\begin{eqnarray}
g_n(N)&=&\frac{g_n(N=2)}{k} \label{eq001} \\
g_l(N)&=&\frac{g_l(N=2) |\gamma_2(N=2)|}{|\gamma_w(N)|} \label{eq002}
\end{eqnarray}

\noindent
Denote $\lambda^{max}_1(N=2)$ and $\lambda^{max}_2(N=2)$ as the
maximal values of $\lambda_1(N=2,g_n)$ and $\lambda_2(N=2,g_n,g_l)$
with respect to $g_n$ and $g_l$.

As an example of how to use Eqs. (\ref{eq000}), (\ref{eq001}) and
(\ref{eq002}) in order to calculate the upper bound for the rate of
information produced in the network, we consider that the neurons in
the network with $N$ nodes are coupled via electrical and excitatory
chemical synapses in an all-to-all configuration (topology II), then
$k=N-1$, $|\gamma_w(N)|=N$ and $|\gamma_2(N=2)|=2$.

Now, we search for a synapse strength range for which two mutually
coupled neurons have an UPPER character.  For example, let us say the
range $g_l(N=2) \in [0,1]$ and $g_n(N=2) \in [2,10]$, in Fig.
\ref{fig_HKS_HC00_19}(E), for two inhibitory mutually coupled neurons. 

From Eqs. (\ref{eq001}) and (\ref{eq002}), as long as the network with
$N$ nodes has $g_n(N) \leq \frac{1}{2(N-1)}$ and $\frac{0.3}{k}\leq
g_l(N) \leq \frac{1}{N}$, then
$\lambda^{max}_1(N)=\lambda^{max}_1(N=2)$ and
$\lambda^{max}_w(N)=\lambda^{max}_2(N=2)$, and therefore for this
synapse range, the maximum of $H_C$ is

\begin{equation}
\max_{g_n,g_l}{[H_C(N,g_n,g_l)]} = \lambda^{max}_1(N=2) + (N-1)\lambda^{max}_2(N=2)
\label{upper_network}
\end{equation}
\noindent
Notice that Eq. (\ref{upper_network}) is valid to any network topology
as long as Eqs. (\ref{neurons04}) and (\ref{neurons07}) can be rescaled.

For very large networks that are very well connected, $g_l(N)$ and
$g_n(N)$ will be very small, since $k$ and $N$ are large. As a
consequence, $\lambda^{max}_1 \approxeq \lambda^{max}_2$, since
neurons are equal, and we can write

\begin{equation}
\max_{g_n,g_l}{[H_C(N,g_n,g_l)]} = N\lambda^{max}_2(N=2)
\label{upper_network1}
\end{equation}
\noindent
which means that the rate of information produced by large UPPER
neural networks whose neurons are highly connected has an upper bound
that increases linearly with the number of neurons. A similar result
is obtained when the neurons are connected with only electrical
synapses \cite{baptista_NJP2008}.

\section{Conclusion}\label{conclusions}

We have studied the combined action of chemical and electrical
synapses in small networks of Hindmarsh-Rose (HR) neurons in
the process of synchronisation and on the rate of information
production.

There are mainly two scenarios for the appearance of complete
synchronisation for the studied inhibitory networks. If the chemical
synapse strength is small, the larger the chemical synapse strength
used the larger the electrical synapse strength needs to be to achieve
complete synchronisation.  Otherwise, if the chemical synapse strength
is large, complete synchronisation appears if the electrical synapse
strength is larger than a certain value.  In the studied excitatory
networks both synapses work in a constructive way to promote complete
synchronisation: the larger the chemical synapse strength is the
smaller the electrical synapse strength needs to be to achieve complete
synchronisation.

When neurons connect simultaneously by electrical and chemical ways,
there is an optimal range of synaptic strengths for which the
production of information is large. For strengths larger than values
within this optimal range, the larger the electrical and chemical
synaptic strengths are the smaller the production of information of
coupled neurons.

In the absence of complete synchronisation, it is intuitive to expect
that excitatory networks have neurons that are more desynchronous while
inhibitory networks have neurons that are more synchronous. This intuitive
idea can be better formalised by understanding the relationship
between excitation (inhibition), synchronisation (desynchronisation)
and the rate of information production. For that we classify the
network as having an UPPER or a LOWER character. In a UPPER (LOWER)
network, the sum of all the positive Lyapunov exponents, denoted by
$H_L$, is bounded from above (below) by the sum of all the positive
conditional Lyapunov exponents, denoted by $H_C$, the Lyapunov
exponents of the synchronisation manifold and the transversal
directions. Networks that have neurons connected simultaneously by
inhibitory chemical synapses and electrical synapses can be expected
to have an UPPER character.  In such networks, one should expect to
find synchronous behaviour, since the synapses force the trajectory
to approach the synchronisation manifold.  On the other hand, networks
whose chemical synapse are of the excitatory type might likely have a
LOWER character. In such networks one should expect to find
desynchronous behaviour since the synapses force the trajectory to
depart from the synchronisation manifold.

Notice that $H_L(N)$ can only be numerically obtained whereas $H_C(N)$
can be calculated from the conditional exponents numerically obtained
for two mutually coupled neurons that have equal trajectories.  For
UPPER networks, $H_C(N) > H_L(N)$, and by Ruelle \cite{ruelle} $H_L(N)
\geq H_{KS}(N)$, where $H_{KS}$ is the Kolmogorov-Sinai entropy, the
amount of information (Shannon's entropy) produced by time unit;
we have then that $H_C$ is an upper bound for $H_{KS}(N)$. That can be
advantageously used in order to calculate the rate of information
produced by a large network, composed of $N$ neurons by using only the
rate at which information is produced in two mutually coupled neurons
that are completely synchronous and have equal trajectories.

We have worked with idealistic networks. However, our results can be
extended to more realistic networks \cite{Baptista1}. For UPPER
networks, our numerical results show that more realistic networks
constructed with non-equal nodes (or networks of equal nodes but with
random synapse strengths \cite{baptista_PLOS2008}) have $H_L$ smaller
than the networks with equal nodes. Therefore, even though networks
with equal nodes might not be realistic, their entropy production per
time unit is an upper bound for the entropy production of more
realistic networks.

\textbf{Acknowledgment} MSB and FMMK thank the Max-Planck-Institut
f\"ur Physik komplexer Systeme (Dresden) for the partial support of
this research.  MSB acknowledges the partial financial support of
"Funda\c c\~ao para a Ci\^encia e Tecnologia (FCT), Portugal" through
the programmes POCTI and POSI, with Portuguese and European Community
structural funds. The authors are deeply grateful for the 4
anonimous referees for their important comments and suggestions that
were considered in this new version of the manuscript.

\section{Appendix}\label{apendice1}

\subsection{A lower bound for the KS entropy}\label{lower_saco}

Imagine a 2D chaotic system as the one studied in Ref.
\cite{baptista_PRE2008} [Eqs. (5) and (6)]. Following the same ideas
from there, the KS entropy of two coupled maps with variables
$x^{\alpha}$ and $x^{\beta}$ can be estimated from the Shannon's
entropy of the probabilities that a trajectory point makes a given
itinerary in the phase space $(x^{\alpha},x^{\beta})$, divided by the
time interval for the trajectory to make that itinerary. 

In practice, calculating the Shannon's entropy \cite{shannon} for all
possible itineraries on the phase space ($x^{\alpha}$,$x^{\beta}$) of
a chaotic trajectory is equivalent to calculating the joint entropy
between the probabilities of finding a point following simultaneously
an itinerary along the variable the variable $x^{\alpha}$ and another
itinerary along the variable $x^{\beta}$.

Since we are unable to make a high resolution partition of the phase
space (nor we do not know the Markov partition) in the neural networks
studied in this work, we estimate a lower bound for the KS entropy by
calculating the joint entropy between symbolic sequences encoding the
trajectory.  Such calculation of probabilities involve large matrix
operations and for that reason we restrain ourselves to the
calculation of the joint entropy between two neurons.

It is a lower bound due to two reasons. The first one is because the
entropy will be measured considering the probabilities of occupation
of a projected trajectory in a subspace of the network. The second one
is because we calculate the entropy considering the probabilities of
binary symbolic sequences and obviously a binary sequence may contain
much less information than the content of a continuous signal
\cite{paninski}.

In the following, we show in more details how this estimation is done.
The way we encode the trajectory is partially based on the time
encoding proposed in Ref.  \cite{baptista_PLOS2008}.

Given two symbolic sequences $S_1$ and $S_2$, 
generated by neuron 1 and 2, respectively, a lower bound for the KS entropy can be estimated by 
\begin{equation}
  H_{low} = \frac{1}{\langle \tau \rangle} H(S_1;S_2)
\label{shannon1}
\end{equation}
with $H(S_1;S_2)$ representing the joint entropy between the symbolic
sequences $S_1$ and $S_2$.  To create the symbolic sequences, we
represent the time at which the $n$-th maxima happens in neuron 1 by
$T_1^n$, and the time interval between the n-th and the (n+1)-th
maxima, by $\delta T_1^n$. A maxima represents the moment when the
action potential reaches its maximal value.  The quantity $\langle
\tau \rangle$ represents the average time between two spikes.  We then
encode the spiking events using the following rule.  The $i$-th symbol
of the encoding is a ``1'' if a spike is found in the time interval
$[i\Delta, (i+1) \Delta[$, and ``0'' otherwise.  We choose $\Delta \in
[\min{(\delta T_1^n)}, \max{(\delta T_1^n)}]$ in order to maximise
$H_{low}$. Each neuron produces a symbolic sequence that is split
into small non-overlapping sequences of length $L$=8.

%

%
\end{document}